\documentclass[floatfix,superscriptaddress,a4paper,
               nofootinbib,preprint]{revtex4}

\pdfoutput=1

\usepackage{graphicx}
\usepackage{amsmath}
\usepackage{amssymb}

\newcommand{\hsp}{\hspace*{1pt}}
\newcommand{\hspm}{\hspace*{.5pt}}
\newcommand{\ds}{\displaystyle}
\newcommand{\be}{\begin{equation}}
\newcommand{\ee}{\end{equation}}
\newcommand{\bel}[1]{\be\label{#1}}
\newcommand{\re}[1]{Eq.~(\ref{#1})}

\usepackage{color}

\begin{document}

\title{Bose-Einstein condensation and \\liquid-gas  phase transition in $\alpha$-matter
}

\author{L.~M. Satarov}
\affiliation{
Frankfurt Institute for Advanced Studies, D-60438 Frankfurt am Main, Germany}
\affiliation{
National Research Center ''Kurchatov Institute'' 123182 Moscow, Russia}

\author{M.~I. Gorenstein}
\affiliation{
Frankfurt Institute for Advanced Studies, D-60438 Frankfurt am Main, Germany}
\affiliation{
Bogolyubov Institute for Theoretical Physics, 03680 Kiev, Ukraine}

\author{A.~Motornenko}
\affiliation{
Frankfurt Institute for Advanced Studies, D-60438 Frankfurt am Main, Germany}
\affiliation{
Department of Physics, Taras Shevchenko National University of Kiev, 03022 Kiev, Ukraine}
\affiliation{
Department of Physics, University of Oslo, 0313 Oslo, Norway}

\author{V. Vovchenko}
\affiliation{
Frankfurt Institute for Advanced Studies, D-60438 Frankfurt am Main, Germany}
\affiliation{
Institut f\"ur Theoretische Physik,
Goethe Universit\"at Frankfurt, D-60438 Frankfurt am Main, Germany}
\affiliation{
Department of Physics, Taras Shevchenko National University of Kiev, 03022 Kiev, Ukraine}

\author{I. N. Mishustin}
\affiliation{
Frankfurt Institute for Advanced Studies, D-60438 Frankfurt am Main, Germany}
\affiliation{
National Research Center ''Kurchatov Institute'' 123182 Moscow, Russia}

\author{H. Stoecker}
\affiliation{
Frankfurt Institute for Advanced Studies,
D-60438 Frankfurt am Main, Germany}
\affiliation{
Institut f\"ur Theoretische Physik,
Goethe Universit\"at Frankfurt, D-60438 Frankfurt am Main, Germany}
\affiliation{
GSI Helmholtzzentrum f\"ur Schwerionenforschung GmbH, D-64291 Darmstadt, Germany}

\begin{abstract}
Systems of Bose particles with both repulsive and attractive interactions are studied
using the Skyrme-like mean-field model. The phase diagram of such  systems
exhibits two special lines in the chemical potential\hsp --\hsp temperature plane:
one line which represents the first-order liquid-gas phase transition with the critical
end point, and another line which represents the onset of Bose-Einstein condensation. The calculations are made for strongly-interacting matter composed of $\alpha$ particles. The phase diagram of this matter is qualitatively similar to that observed for the atomic $^4$He liquid. The sensitivity of the results to the model parameters is studied.
For weak interaction coupling the critical point is located at
the Bose-condensation line.
\end{abstract}

\maketitle

\section{Introduction}
Bose-Einstein condensation (BEC) in non-interacting boson~systems
was theoretically predicted a long time ago \cite{Bose,Einstein}.
However, only in 1995 two groups succeeded to create the necessary
experimental conditions of very cold and dilute atomic Bose gases
by using novel developments in cooling and trapping techniques~\cite{BEC1,BEC2}.
Currently,
investigation of quantum phenomena in cold systems of interacting bosons
is a subject of active theoretical~\cite{Pet02,Pit03} and experimental~\cite{Chi10}  studies.

The first-order liquid-gas phase transition (LGPT), on the other hand, is a well-known
phenomenon that occurs in systems of interacting atoms or molecules~\cite{Lan75,Gre97}
as well as in nuclear matter~\cite{Kue74,Jaq83}.
The LGPT is a consequence of inter-particle interactions containing both attractive and repulsive forces. The observed phase diagram of atomic $^4$He~\cite{Buc90} contains simultaneously
the regions of the LGPT and the BEC.
Note that the superfluidity phenomenon in Bose liquids appears only for nonzero
interactions of particles~\cite{Lif80}.

In the present paper we study the phase
diagram of matter composed of~$\alpha$~particles with strong (nuclear) interactions.
It is commonly accepted that $\alpha$ particles and $\alpha$-cluster correlations are important in atomic nuclei, in intermediate-energy heavy-ion collisions,
and in astrophysical environments.
This is due to the~large binding energy of $\alpha$ particles as compared to
other light clusters like $d,t,^3$He.
Theoretical and experimental
studies of nuclear systems with $\alpha$ clusters have already a~long history. The energy
density of cold homogeneous $\alpha$-matter was estimated by using realistic
$\alpha\hspm\alpha$~potentials in Refs.~\cite{Cla66,Bri73}. It was shown that
such $\alpha$-matter is energetically
favorable at low baryon densities (i.e., it
has a larger binding energy per baryon) as compared to uniform nucleon matter.

It is believed that $\alpha$ clusters exist either at periphery of heavy nuclei or in low-density excited states of light nuclei like $^{12}$C, $^{16}$O, $^{20}$Ne,
etc.~\cite{Oer06,Bec12,Sch16}. Especially interesting is the second excited (Hoyle) state of $^{12}$C which plays a key role in stellar nucleosynthesis. As argued in Ref.~\cite{Roe98} the Hoyle state can be regarded
as the Bose-condensed coherent state of~3\hsp$\alpha$ particles. Enhanced yields of~$\alpha$ particles have been observed in multifragmentation reactions in heavy-ion colli\-sions~\mbox{\cite{Bon95,Mar16,Bor16}}.
Considerable abundances of~$\alpha$ clusters are expected in the outer regions of compact
stars, in neutron star mergers, and in supernova mat\-ter~\cite{Typ10,Bot10}.

Different theoretical methods were used to study the equation of state of
nuclear matter with nucleons and light nuclei at various temperatures and baryon densities.
In particular, phenomenological liquid-drop models were applied in Refs.~\cite{Lat91,Buy14}.
The chemically equilibrated $N+\alpha$ mixture has been studied~\cite{Hor06} by using the virial expansion, and the relativistic mean-field approach was used in Refs.~\cite{Typ10,Mis17}.
Lattice calculations for $\alpha$-matter were presented in Ref.~\cite{Sed06}.

In the present study we  calculate the phase diagram of pure $\alpha$-matter using the
mean-field Skyrme-like interaction. In contrast to other authors, we study simultaneously both, the LGPT and~BEC. The article is organized as follows. In Sec.~II the  model is formulated.
The phase diagrams for different pairs of thermodynamical variables
are presented and discussed in Sec.~III. The sensitivity of the results to the
model parameters is also considered in this section.
Conclusions and outlook are given in Sec.~IV.

\section{Bosonic system with~Skyrme interaction}

Let us consider a system of interacting bosons with mass $m$ and degeneracy
factor $g$. The grand canonical pressure $p\hspm (T,\mu)$ is a function of tempera\-ture~$T$ and chemical potential $\mu$ and plays the role of the ther\-mo\-dynamical potential. The particle number
density~$n\hspm (T,\mu)$, the entropy density $s\hspm (T,\mu)$, and the energy density $\varepsilon\hspm (T,\mu)$ are calculated from~$p\hspm (T,\mu)$ as
\bel{therm}
n=\left(\frac{\partial p}{\partial \mu}\right)_T,~~~~ s=\left(\frac{\partial p}{\partial T}\right)_{\ds\mu},
~~~~\varepsilon=T s+\mu\hsp n-p
\ee
in the thermodynamic limit, where the system volume goes to infinity.

In the present paper,
both attractive and repulsive interactions are described in the mean-field
approximation, by introducing the potential $U(n)$ which depends on the
particle density~$n$, but does not depend on $T$~\cite{Hof76,Gal78}.
This is achieved by shifting the chemical potential with respect to the ideal gas value,
\bel{mu6}
\mu^*~=~\mu~-~U(n)\hsp .
\ee
Thermodynamical consistency requires then an additional (field) term in the pressure
$p_f$ (for more details see Refs.~\cite{Mis91,Sat09}):
\bel{pf}
p\hspm (T,\mu)~=~p_{\hspm\rm id}\hspm (T,\mu^*)~+~p_f\hspm (n)\hsp ,
~~~~~p_f\hspm (n)~=~n\,U\hspm (n)~-~\int\limits_0^ndn'\hsp U\hspm (n')\hsp .
\ee
In general, particle number density can be written as
\bel{n1}
n\hspm (T,\mu)~=~n_{\hsp\rm id}(T,\mu^*)~+~n_{\rm bc}~.
\ee
The second term denotes the density of
the Bose condensate which consists of zero-momentum particles.
An explicit procedure of calculating $n_{\hsp\rm bc}$ is given below.

The ideal Bose gas pressure and particle number density in Eqs.~(\ref{pf}) and (\ref{n1})
are, respectively, ($\hbar=c=1$)
\begin{eqnarray}
&&p_{\hsp\rm id}(T,\mu^*) = \dfrac{g}{6\pi^2}\int\limits_0^{\infty} dk~
\frac{k^4}{\sqrt{k^2+m^2}}\left[\exp{\left(\dfrac{\sqrt{k^2+m^2}-\mu^*}{T}\right)}
-1\right]^{-1},\label{p-id}\\
&&n_{\hsp\rm id}(T,\mu^*) = \dfrac{g}{2\pi^2}\int\limits_0^\infty dk~k^2
\left[\exp \left(\dfrac{\sqrt{k^2+m^2}-\mu^*}{T}\right)-1\right]^{-1}. \label{n-id}
\end{eqnarray}
Note that Eqs.~(\ref{p-id}) and (\ref{n-id}) are only valid for $\mu^* \leqslant m$. A macroscopic fraction of zero momentum particles is accumulated at $\mu^*=m$ forming a Bose condensate with nonzero
density~$n_{\rm bc}$.

In the spirit of the Skyrme model \cite{Ben03,Sto07} we parameterize the mean-field potential as
\bel{skyrme}
U(n)~=~-\hspm a\hspm n \left[2~-~\frac{\gamma+2}{\gamma+1}\,\left(\frac{n}{n_0}\right)^{\gamma}\,\right].
\ee
Here $a$, $n_0$, and $\gamma$ are the positive model parameters.
Substituting~(\ref{skyrme}) into \re{pf} yields
\bel{pf1}
p_f(n)~=~-~a\,n^2 \left[1~-~\left(\frac{n}{n_0}\right)^{\gamma}\,\right]\hsp .
\ee
Note that the first terms on the right hand sides
of Eqs.~(\ref{skyrme}) and (\ref{pf1})
correspond to attractive interactions between particles. Their form is identical to
the van der Waals model \cite{Lan75,Gre97}. The second terms in these equations describe the repulsive interactions at short inter-particle distances.
Note that $p_f(n_0)=0$, and the point $T=0$ and $n=n_0$ defines the ground
state of the system. This is the case for all values $a>0$.
It is a consequence of a special form~(\ref{skyrme}) of the potential
$U(n)$, where both attractive and repulsive terms
are proportional to the coupling constant $a$.

Similar parameterizations of the nucleon mean-field interaction are often used
to  describe properties of cold nucleon matter. Reasonable values of the nuclear matter
compressibility are obtained for $1/6\leqslant\gamma\leqslant 1/3$~\cite{Ben03}. We use the same interval of~$\gamma$ values in our calculations.

The BEC in the ideal Bose gas (i.e., at $a=0$) occurs at $\mu^*=\mu=m$.
At $a>0$, this condition is replaced by $\mu^*=\mu-U\hspm (n)=m$,
which can be written as
\bel{mueq}
\mu ~=~ m~+~U\big(n_{\hsp\rm id}(T,m) + n_{\hsp\rm bc}\big).
\ee
Equation (\ref{mueq}) with $n_{\rm bc}=0$ defines the line $T=T_{\rm BEC}\hsp (\mu)$
in the $(\mu,T)$ plane  which corresponds to the onset of the BEC.
Below this line, Eq.~(\ref{mueq}) gives nonzero values of $n_{\rm bc}$
as a function of~$T$ and $\mu$.

Substituting $\mu^*=m$ into \re{n-id}, one obtains in the lowest order in~$T/m$
\bel{nid1}
n_{\hspm\rm id}\hspm (T,m)~=~g\,\left(\frac{mT}{2\pi}\right)^{3/2}
~\left[\zeta\hsp (3/2)+\frac{\ds 15}{\ds 8}\hsp
\frac{\ds T}{\ds m}\hsp\zeta\hsp (5/2)+\ldots\right].
\ee
Here
$\zeta(x)=\sum_{k=1}^{\infty} k^{-x}$ is the Riemann zeta-function, $\zeta(3/2)\cong 2.612$ and $\zeta(5/2)\cong 1.341$.
The first term on the right hand side of Eq.~(\ref{nid1}) is the well-known nonrelativistic
result~\cite{Lan75,Gre97}, and the second term gives a relativistic correction.
The line $T=T_{\rm BEC}(n)$ obtained from Eq.~(\ref{nid1}) with $n_{\rm id}=n$ describes the onset of the BEC in the $(n,T)$ plane. This line
is `universal', i.e., it does not depend on the interaction strength, and thus coincides with that for the ideal Bose gas. Neglecting relativistic corrections in ~\re{nid1} yields
\bel{tbec}
T_{\rm BEC}(n)~\simeq ~ \frac{2\pi}{m}\left[\frac{n}{\zeta (3/2)\hsp g}\hsp\right]^{2/3}.
\ee
Equation (\ref{tbec}) remains valid for the whole class of mean-field models with any potential~$U(n)$ which does not depend on $T$.

All thermodynamic functions can be calculated by using the above equations.
The main feature of the LGPT is the appearance of
the mixed phase which consists of `gas' ($n=n_g$) and `liquid' ($n=n_l$) domains with
densities $n_g < n_l$ at fixed $T$.
The mixed-phase boundaries (the so-called binodals) in the $(n,T)$ plane are found from the Gibbs conditions of the phase equilibrium:
\bel{gceq}
p\hspm (n_g,T)~=~p\hspm (n_l,T),~~~~~~\mu\hsp (n_g,T)~=~\mu\hsp (n_l,T)\,.
\ee
The critical point (CP), $T=T_{\rm c}$ and $n=n_{\rm c}$, is the 'end point' of the LGPT line in
the~$(\mu,T)$ plane. At this point, both, the first and the second density derivative of the pressure vanish, $(\partial p/\partial\hspm n)_T=0$ and
$(\partial^{\hsp 2} p/\partial\hspm n^2)_T=0$ \cite{Lan75,Gre97}.

Fluctuations of thermodynamic variables play an important role in the physics of phase transitions. Especially interesting for nuclear systems are fluctuations of
the particle number in a fixed volume. They are characterized by the scaled
vari\-ance of particle number
fluctuations,~\mbox{$\omega=\langle (\Delta N)^2\rangle/\langle N\rangle $}.
 Using Eqs.~(\ref{n-id}) and  (\ref{mu6}), one obtains the following expression~(see also Refs.~\cite{Vov15,Vov17}):
\bel{variance}
\omega~=~{T\left(\frac{\partial p}{\partial\hspm n}\right)^{-1}_T=}~
\frac{T}{n}\left(\frac{\partial\hsp n}{\partial \mu}\right)_T=~\omega_{\hsp\rm id}\hspm (T,\mu^*)
\left[1+\dfrac{n\hsp U^{\hsp\prime}\hspm (n)\,\omega_{\hsp\rm id}\hspm (T,\mu^*)}{T}\right]^{-1}\,,
\ee
where $\omega_{\hsp\rm id}=(T/n_{\hspm\rm id})\,\left(\partial\hsp n_{\hspm\rm id}/
\partial \mu^*\right)_T$ is the scaled variance for the ideal Bose gas.

The condi\-tions~$(\partial\hsp p/\partial\hspm n)_{\hsp T}=0$ and $\omega=\infty$
are fulfilled at the CP (see e.g., Refs.~\cite{Lan75,Gre97}).
This in turn implies that the relation $nU^\prime=-T/\omega_{\rm id}$ holds at the CP.
Note that~\re{variance} is not applicable for unstable states where
$(\partial p/\partial\hspm n)_{\hsp T}<0$.

In addition to the CP, the BEC is another potential source of anomalous particle number
fluctuations: indeed, in the ideal Bose gas one has $\omega_{\rm id}\rightarrow\infty$
at the onset of the BEC \cite{BG}, i.e., at $\mu=\mu^*\rightarrow m$. This is, however, not the case
in the model considered here. Using~\re{variance} one obtains the finite values
$\omega=T/[nU^{\prime}(n)]$ in the limit $\mu^*\rightarrow m$.

\section{Phase diagram of {\Large $\alpha$}--matter}

In this section we present results for the matter composed of $\alpha$-particles
(\mbox{$g=1$}, \mbox{$m\simeq 3727~\textrm{MeV}$}).
Unless stated otherwise, the calculations are done for the following set of the model parameters:
\mbox{$a=3~{\rm GeV~fm^3}$}, \mbox{$\gamma=1/6$}, \mbox{$n_0=0.05~\textrm{fm}^{-3}$}.
The corresponding values for $T_c$ and~$n_c$ are presented in Table~I.

In the Boltzmann approximation, $p_{\hsp\rm id}=nT$ and $\omega_{\hsp\rm id}=1$,
one gets the following simple relations for critical temperature and density:
\bel{crp}
T_{\rm c}~=~-~n_{\rm c}\hsp U^{\hsp\prime}(n_{\rm c})~=~n_{\rm c}^2\hsp
U^{\hsp\prime\prime}(n_{\rm c})\,.
\ee
The second equality follows from $(\partial^{\hsp 2} p/\partial\hspm n^2)_T=0$.
Equations~(\ref{skyrme}) and (\ref{crp}) lead to
\bel{crp1}
T_{\rm c}=\frac{\ds 2\gamma}{\ds \gamma+1}\hsp a\hspm n_{\rm c}~,~~~~~~n_{\rm c}=
n_0\left[\frac{\ds 2}{(\gamma+1)\hspm (\gamma+2)}\right]^{1/\gamma}\,.
\ee
One can see that $T_{\rm c}$ is proportional to $a$, and $n_{\rm c}$ is independent of $a$.
For a given set of model parameters~\re{crp1} one obtains
\mbox{$T_{\rm c}\simeq 10.5~{\rm MeV}$} and
\mbox{$n_{\rm c}\simeq 0.012~{\rm fm^{-3}}$}, in a good agreement with the values
in Table I. However, as will be shown below, the Boltzmann approximation breaks down
for~$a\to 0$\hspm .

\begin{figure}[ht!]
\centering
\includegraphics[trim=1.7cm 7.2cm 2cm 8.5cm, clip, width=0.47\textwidth]{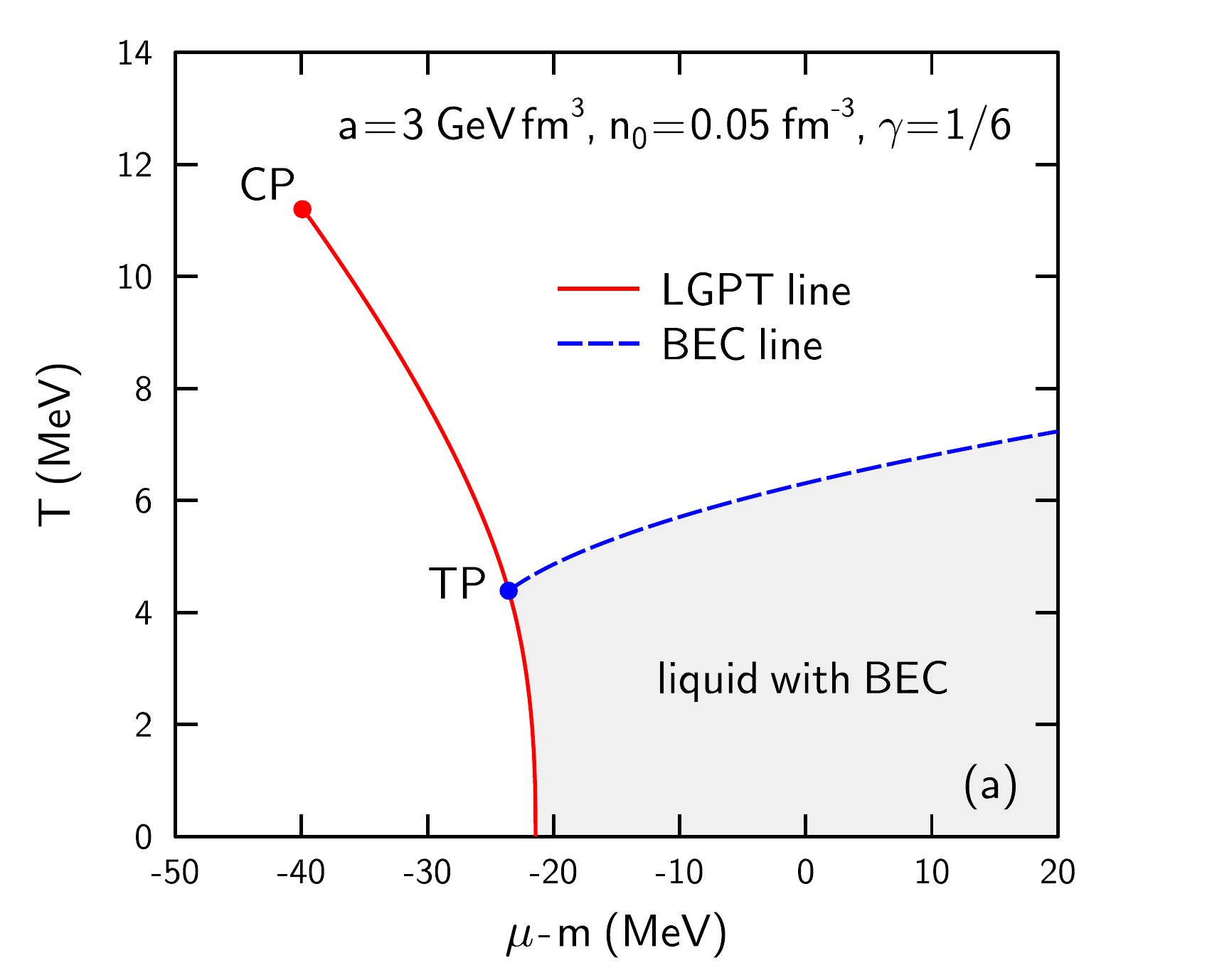}
\includegraphics[trim=1.7cm 7.2cm 2cm 8.5cm, clip, width=0.47\textwidth]{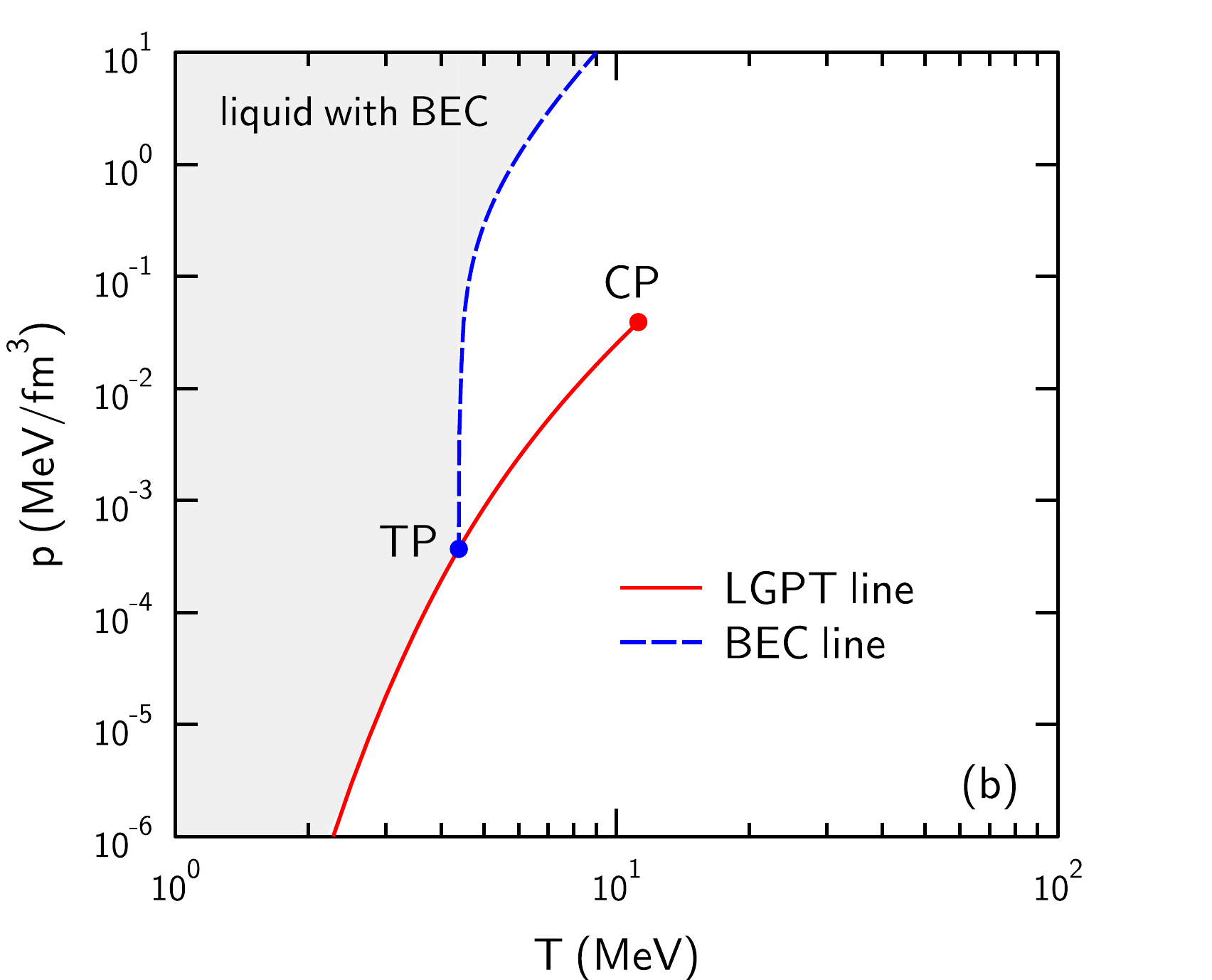}
\caption{
Phase diagram of $\alpha$-matter in $(\mu,T)$ (a) and $(T,p)$ (b) planes. Solid
lines correspond to the LGPT and
dashed lines to the onset of the BEC. The CP and TP
are marked by dots.
}\label{fig-1}
\end{figure}

In Figs.\,\ref{fig-1} (a) and (b) the first-order
LGPT is shown by  solid lines and the BEC by  dashed lines
in the $(\mu,T)$  and $(T,p)$ planes, respectively. The LGPT line corresponds to the
mixed phase states. At $T<T_{\rm c}$, there is
a~discontinuity between the particle densities ($n_l>n_g$) on the two sides of
this line.
The entropy and energy density jump across the LGPT line,
whereas the pressure is continuous.

The shaded regions in Fig.~\ref{fig-1}
correspond to states with nonzero condensate den\-sity~$n_{\rm bc}$.
The CP lies above the~BEC boundary.
Thus, the LGPT and the BEC lines intersect at some point which we call for brevity
as the triple point (TP). Characteristics of the CP and TP are given in Table~I.
The phase diagram of the $\alpha$-matter
in the $(T,p)$ plane shown in Fig.\,\ref{fig-1} (b) is qualitatively similar to
that observed~\cite{Buc90} for atomic $^4$He-matter\hsp\footnote
{
In that case the BEC region and the TP are usually called as the HeII phase and
the $\lambda$ point, respectively.
}.

\begin{table}[ht!]
\caption
{\footnotesize Characteristics of the CP and TP for $\alpha$-matter with
Skyrme interaction (here~\mbox{$a=3~\textrm{GeV\,fm}^3$}, \mbox{$n_0=0.05~\textrm{fm}^{-3}$},
\mbox{$\gamma=1/6$} are used).}
\label{tab1}
\vspace*{3mm}
\begin{tabular}{|c|c|c|c|c|c}\hline
&~$T ~(\textrm{MeV})$~&~$n~(\textrm{fm}^{-3})$~&~$p~(\textrm{MeV/fm}^3)$~&~$\mu-m~(\textrm{MeV})$~\\
\hline
~CP~~&11.2&0.013&$3.92\cdot 10^{-2}$&$-39.9$\\
\hline
TP&4.39&0.045&$3.08\cdot 10^{-4}$&$-23.6$\\
\hline
\end{tabular}
\end{table}

\begin{figure}[ht!]
\centering
\includegraphics[width=0.47\textwidth]{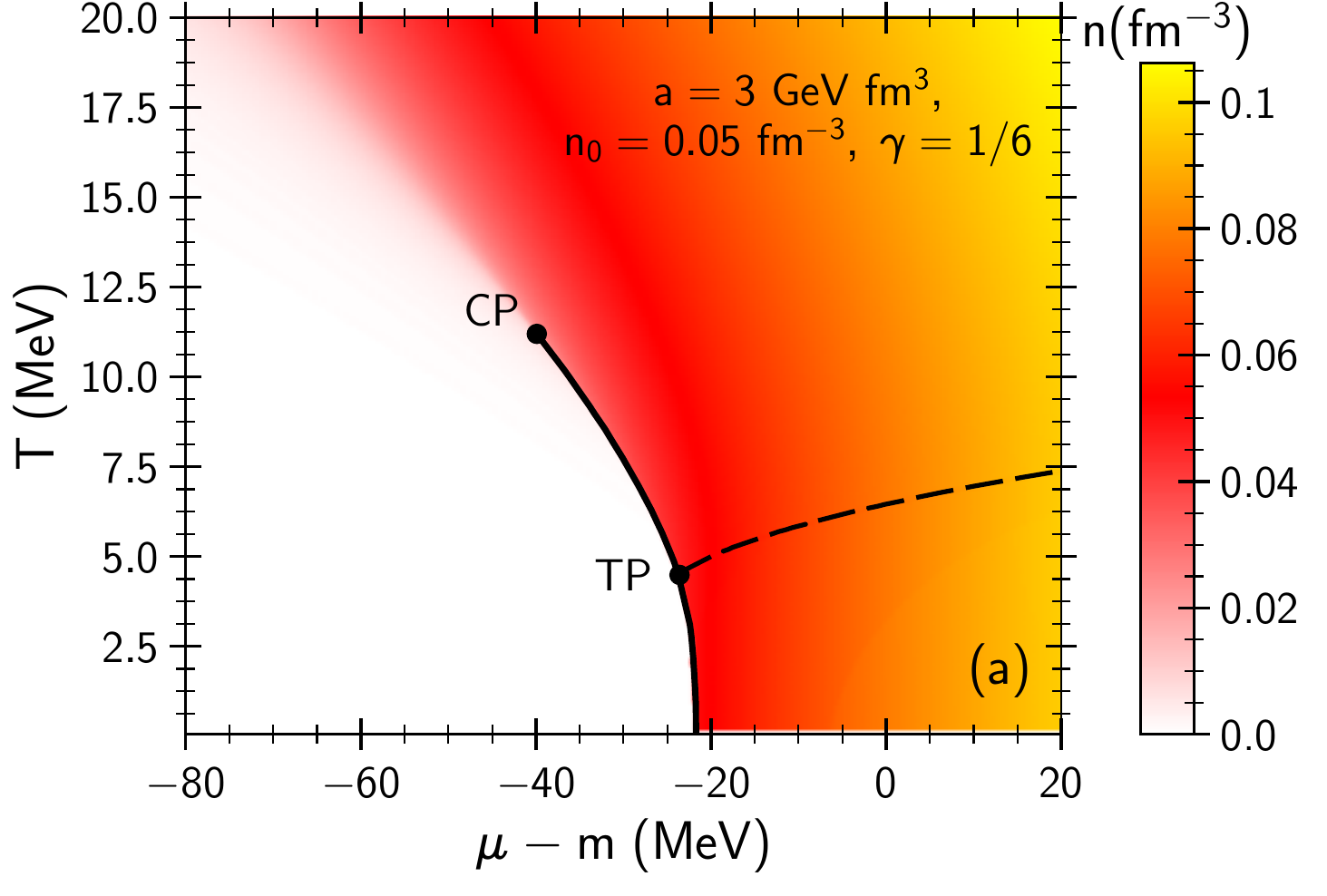}
\includegraphics[width=0.47\textwidth]{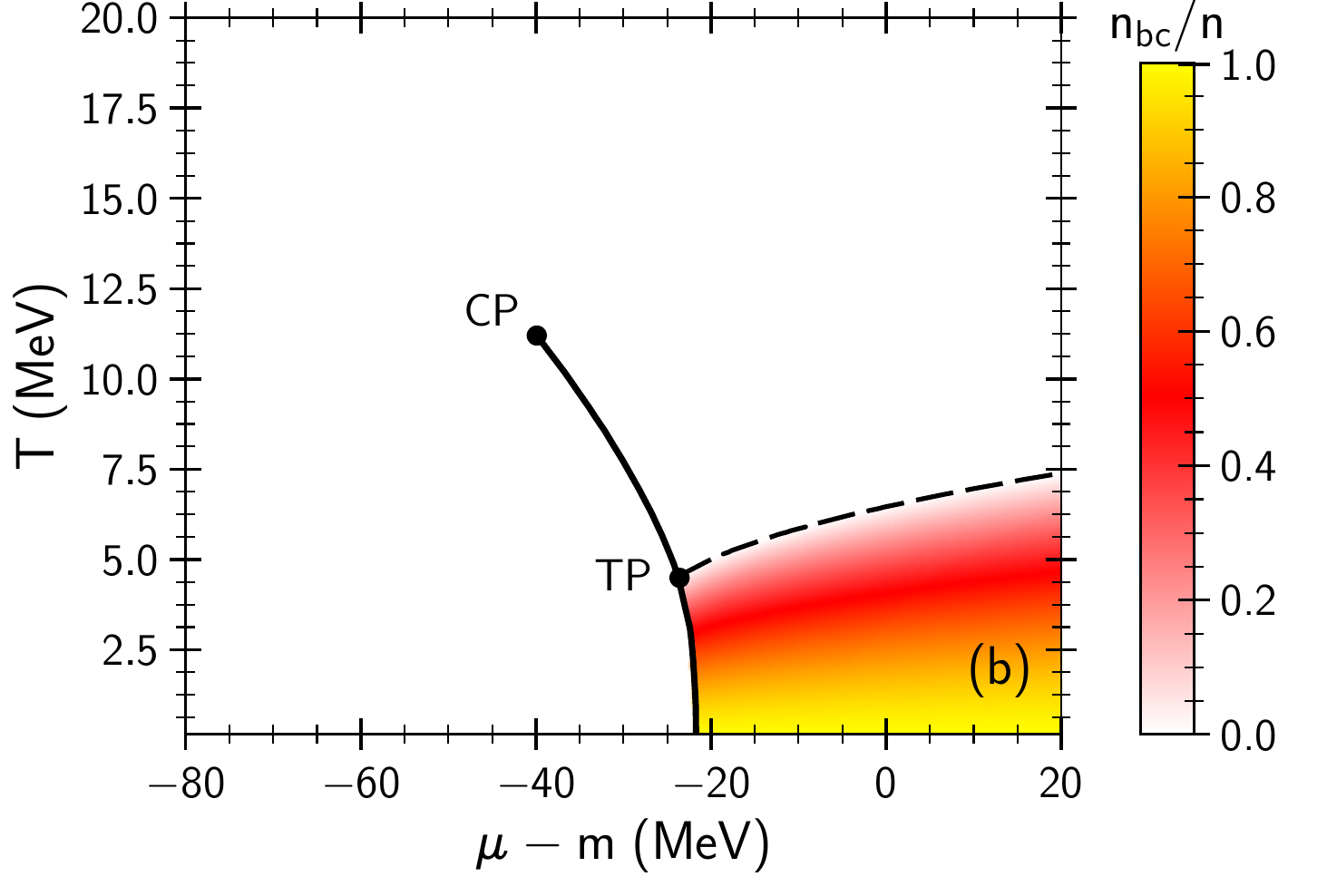}
\includegraphics[width=0.47\textwidth]{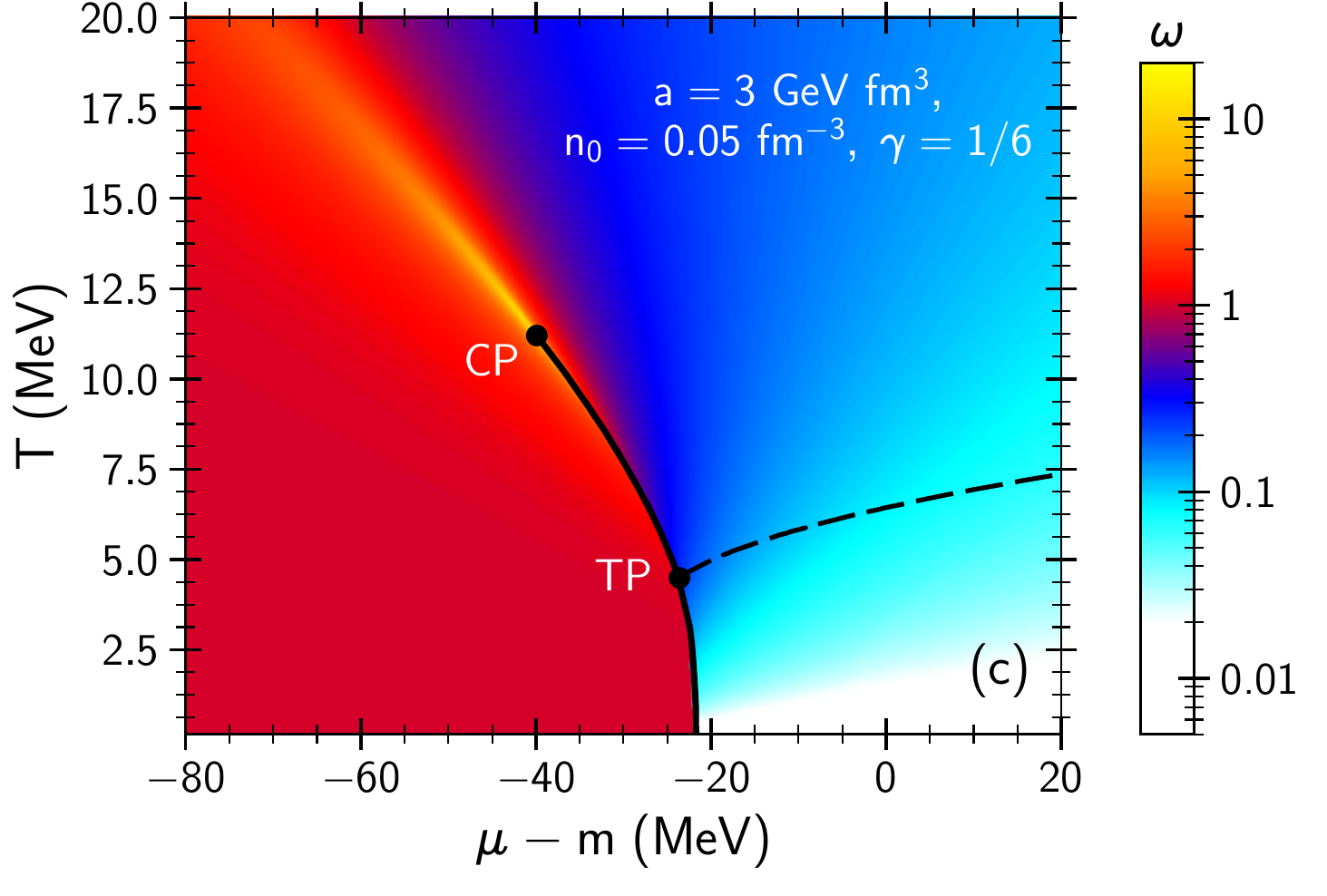}
\includegraphics[width=0.47\textwidth]{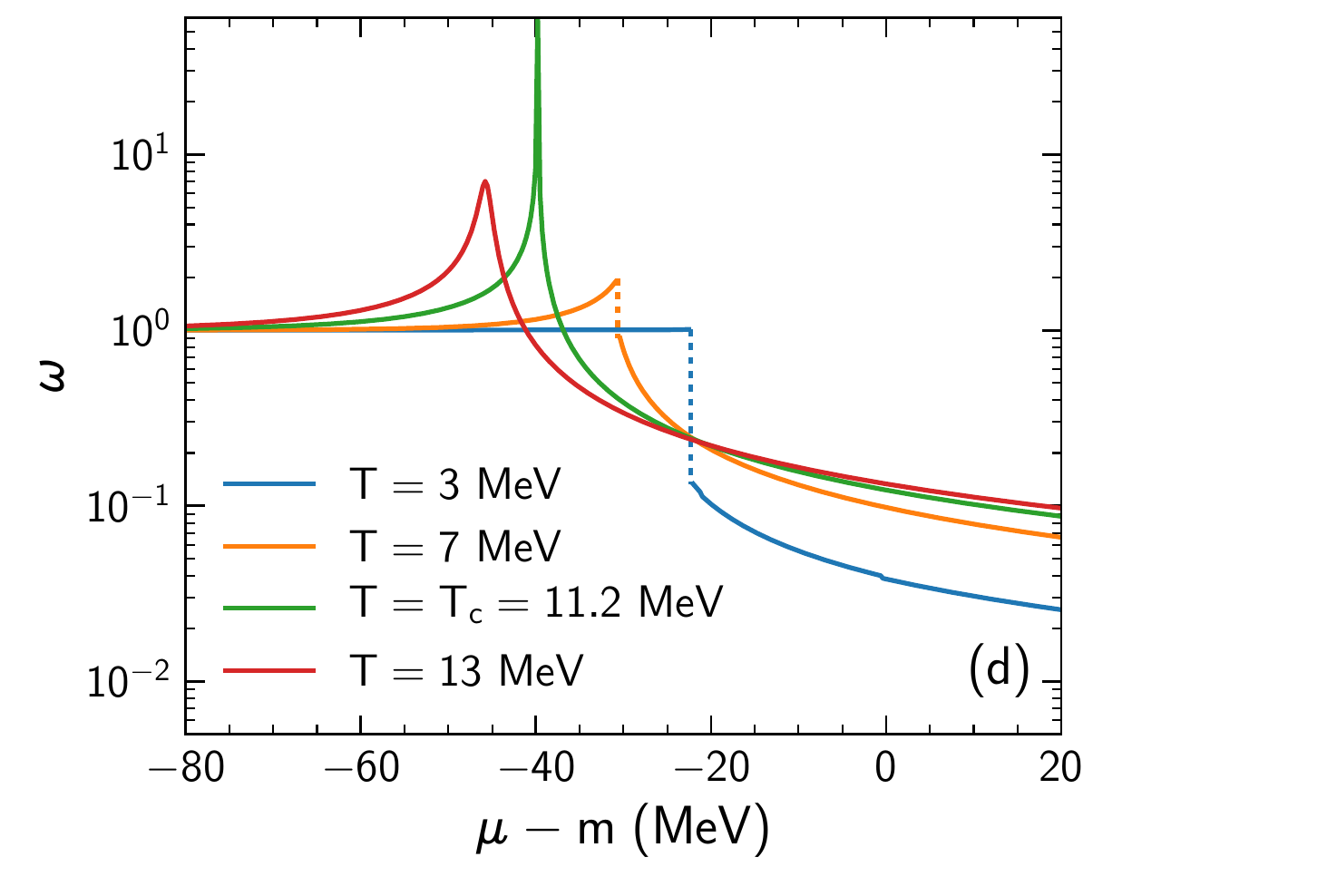}
\caption{Contour plots of the total density $n$ (a), the relative fraction
of Bose-condensed par\-ticles~$n_{\rm bc}/n$~(b),
and the scaled variance~$\omega$ (c) calculated for $\alpha$-matter.
Panel (d) shows isotherms of~$\omega$ for several values of $T$.
The solid and dashed curves show the LGPT and BEC lines, respectively.
The CP and TP are marked by dots.
}\label{fig-2}
\end{figure}
Figures ~\ref{fig-2} (a) and (b)  show, respectively,
the contour plots of the density $n$ and the ratio $n_{\rm bc}/n$ in the $(\mu,T)$ plane.
The LGPT is shown in Fig.~\ref{fig-2} by a solid line.
The discontinuity of the density $n$ across this line is clearly visible
in Fig.~\ref{fig-2} (a).
At $T<T_{\rm TP}$ the gas of $\alpha$-particles with $n=n_{\rm g}$ and $n_{\rm bc}=0$ coexists with the liquid domains with $n=n_{\rm l}>n_{\rm g}$ and nonzero $n_{\rm bc}$.
One can see that above the CP
very rapid, although continuous, changes of density occur
in a narrow region of the $(\mu,T)$ plane. This is a manifestation
of the so-called crossover phenomenon.

The onset of the BEC is shown in Fig.~\ref{fig-2} by dashed lines.
The Bose condensate exists only in the liquid phase below this line.
Figure~\ref{fig-2} (b) shows that the normalized density of
the Bose condensate, $n_{\rm bc}/n$, starts from zero
on the dashed line and goes to its maximum value, $n_{\rm bc}/n=1$, for~$T\rightarrow 0$.

The contour plot of $\omega$ in the $(\mu,T)$ plane is shown in~Fig.~\ref{fig-2} (c).
We have checked that $\omega$ is positive and finite for all states except the CP
where it goes to infinity. Such a behavior strongly deviates from
the ideal Bose gas where $\omega=\omega_{\hsp\rm id}=\infty$
for all states with~$\mu=\mu^*=m$~\cite{BG}. A more detailed information
is shown in Fig.~\ref{fig-2} (d) which represents $\omega$ isotherms for
temperatures above, below and near $T_c$. One can see jumps of $\omega$ across the LGPT
line.

\begin{figure}[ht!]
\centering
\includegraphics[trim=1.7cm 7.5cm 2cm 8.5cm, clip, width=0.47\textwidth]{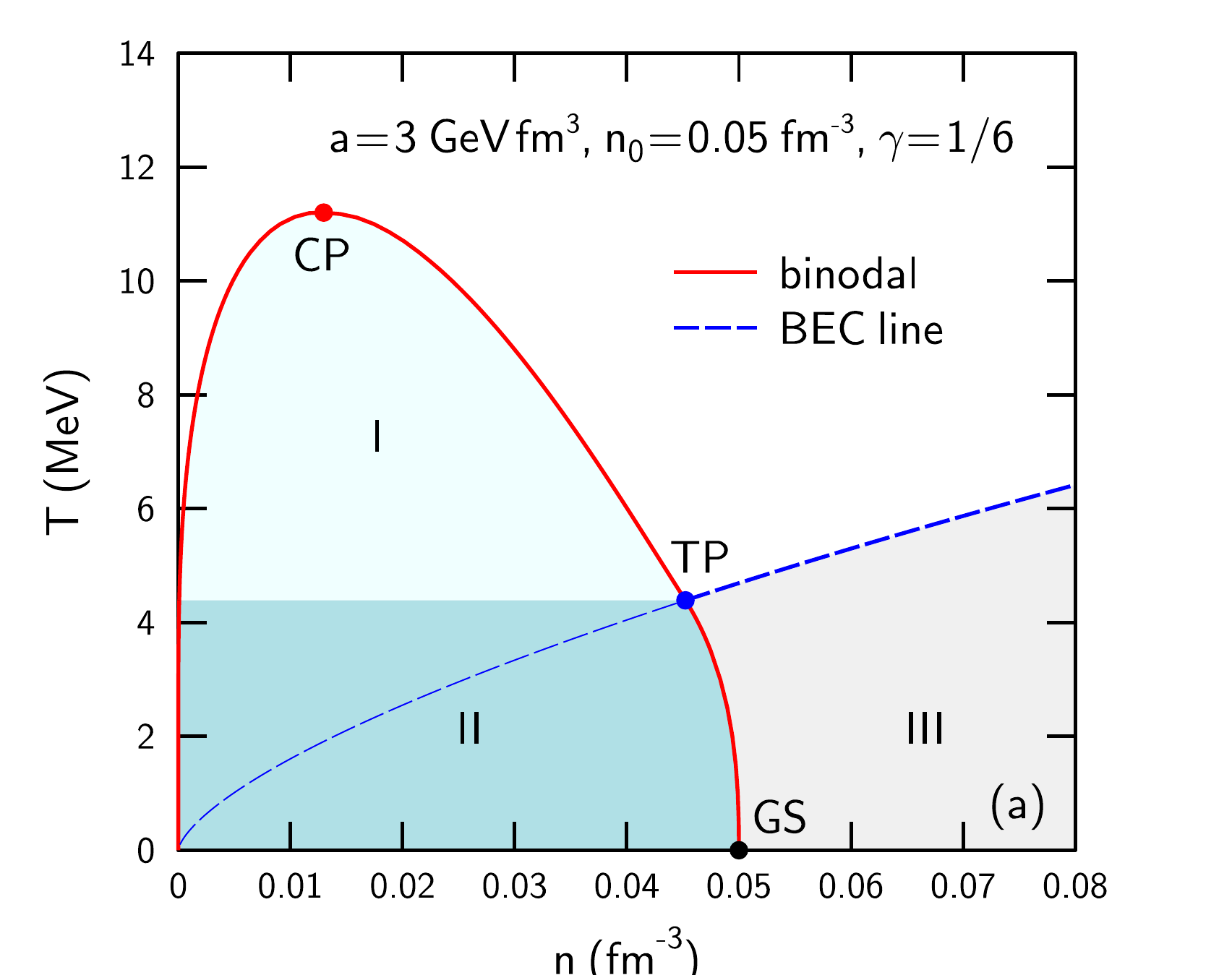}
\includegraphics[trim=1.7cm 7.5cm 2cm 8.5cm, clip, width=0.47\textwidth]{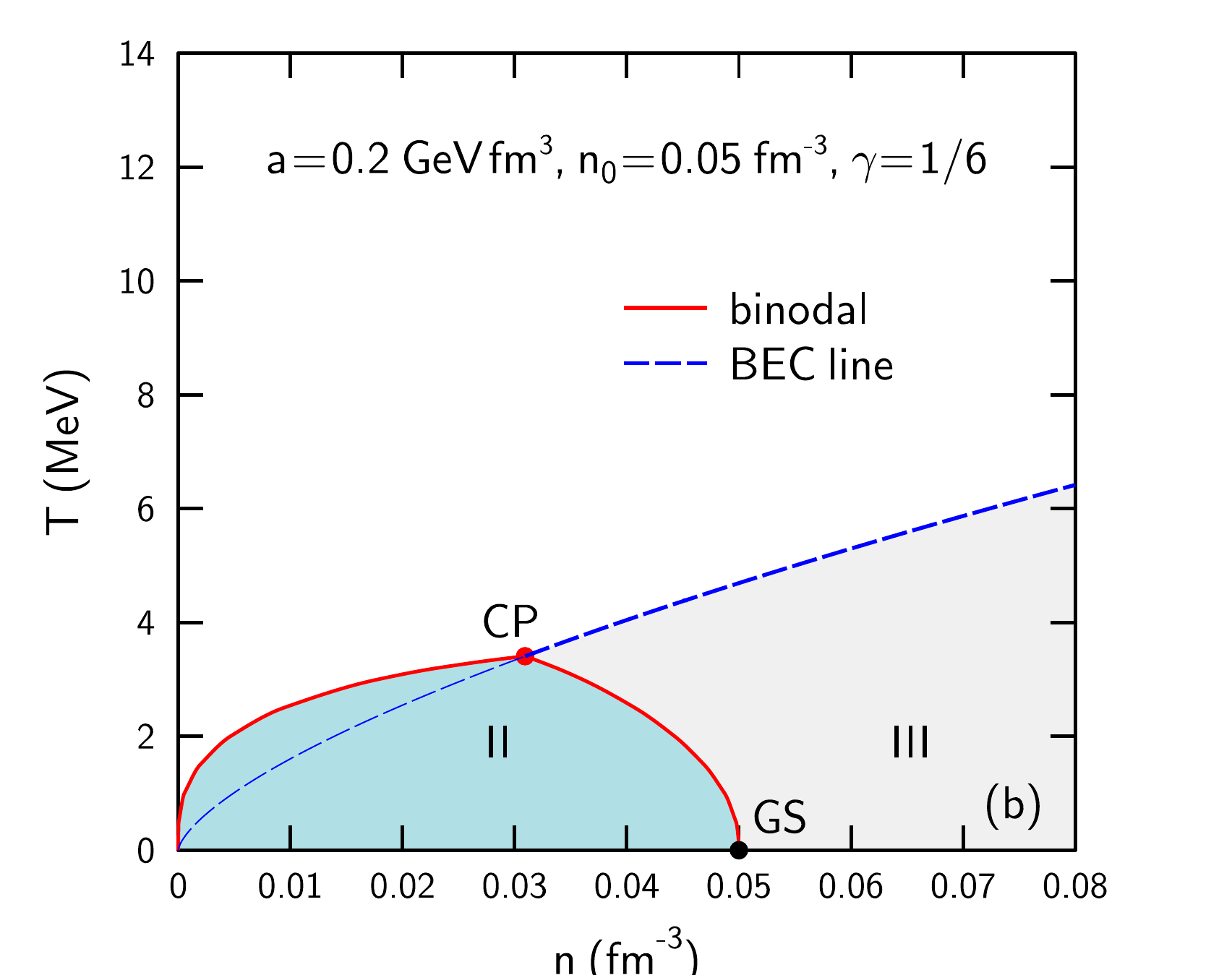}
\includegraphics[trim=1.7cm 7.5cm 2cm 8.5cm, clip, width=0.47\textwidth]{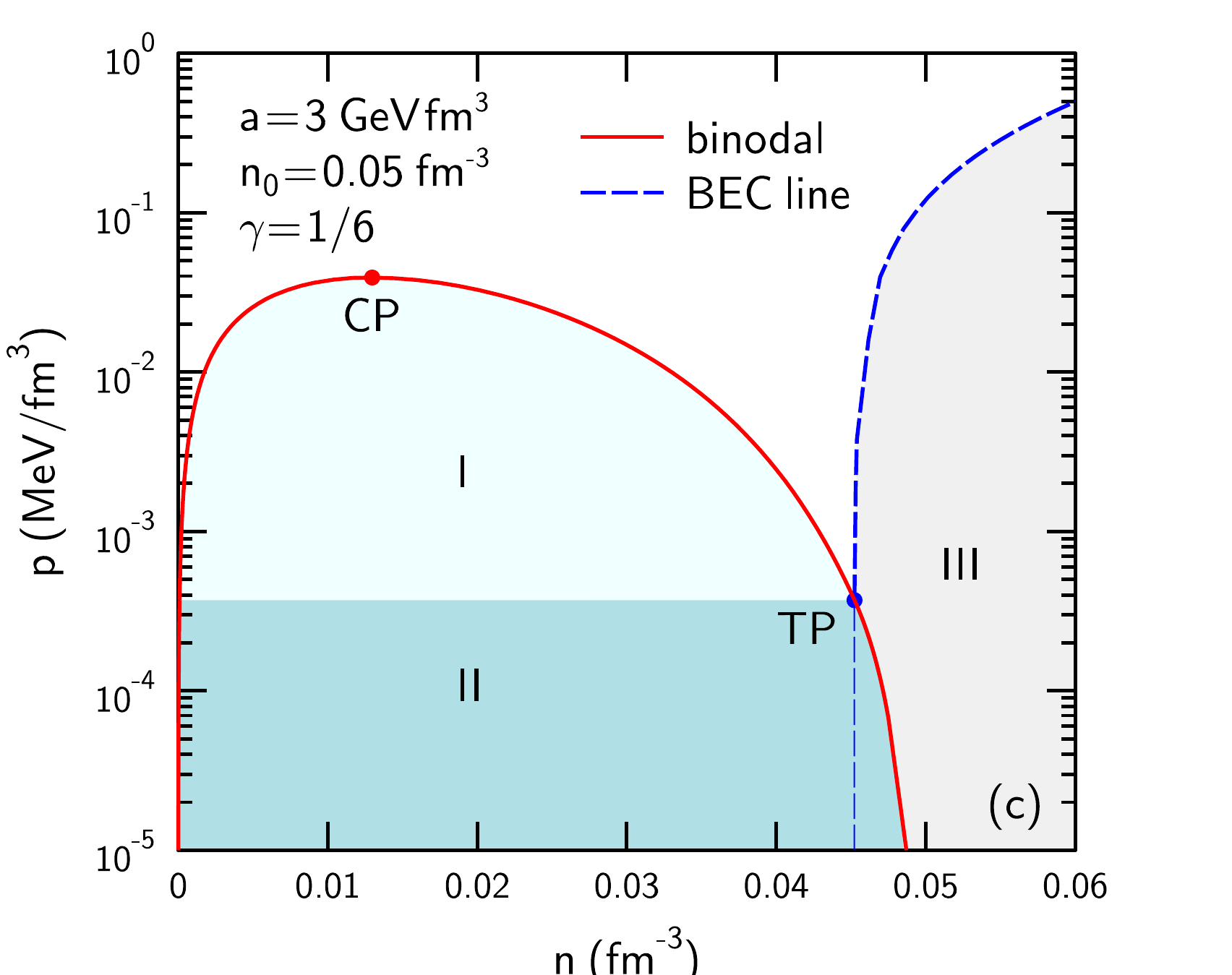}
\includegraphics[trim=1.7cm 7.5cm 2cm 8.5cm, clip, width=0.47\textwidth]{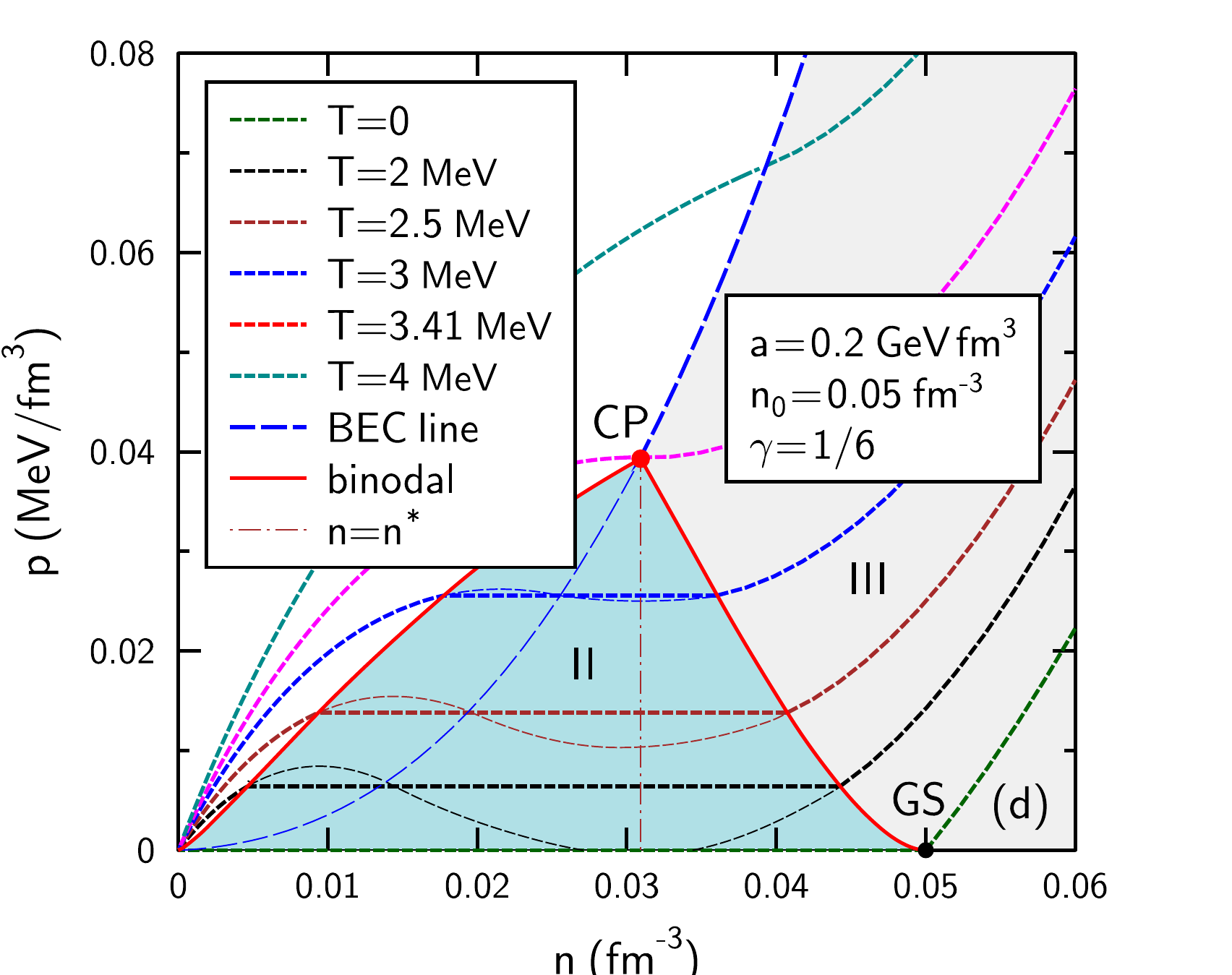}
\caption{Solid lines in all panels show the boundary of the mixed phase, and dashed
curves correspond to the BEC lines.
(a): The phase diagram of the $\alpha$-matter
in the $(n,T)$ plane for the same model parameters as in Figs.~\ref{fig-1} and \ref{fig-2}.
The region I corresponds to the mixed phase with $n_{\rm bc}=0$,
the region II to the mixed phase with nonzero $n_{\rm bc}$
in the liquid component, and the region~III to the pure liquid phase with $n_{\rm bc}>0$.
(b): The same as (a) but for $a=0.2$~GeV~fm$^3$.
Here the~CP and TP coincide, and the region~I is absent (see text for details).
(c): The phase diagram in the~$(n,p)$ plane
for the same parameters as in panel (a).
The logarithmic scale is used to show the region~II.
(d): The same as (c) but for $a=0.2~\textrm{GeV\,fm}^3$ (see text for details).
}
\label{fig-3}
\end{figure}

Figures~\ref{fig-3} (a) and (c) show the phase diagram of the $\alpha$-matter
in the $(n,T)$ and $(n,p)$ planes, respectively,
for the same model parameters as in Fig.~\ref{fig-1}.
The~ground state (GS), $T=0$ and $n=n_0$, has the minimum energy per particle and zero pressure.
Thin dashed curves represent parts of the BEC lines inside the mixed phase region.

\begin{figure}[ht!]
\centering
\includegraphics[trim=2cm 7.5cm 2cm 8.5cm, clip, width=0.47\textwidth]{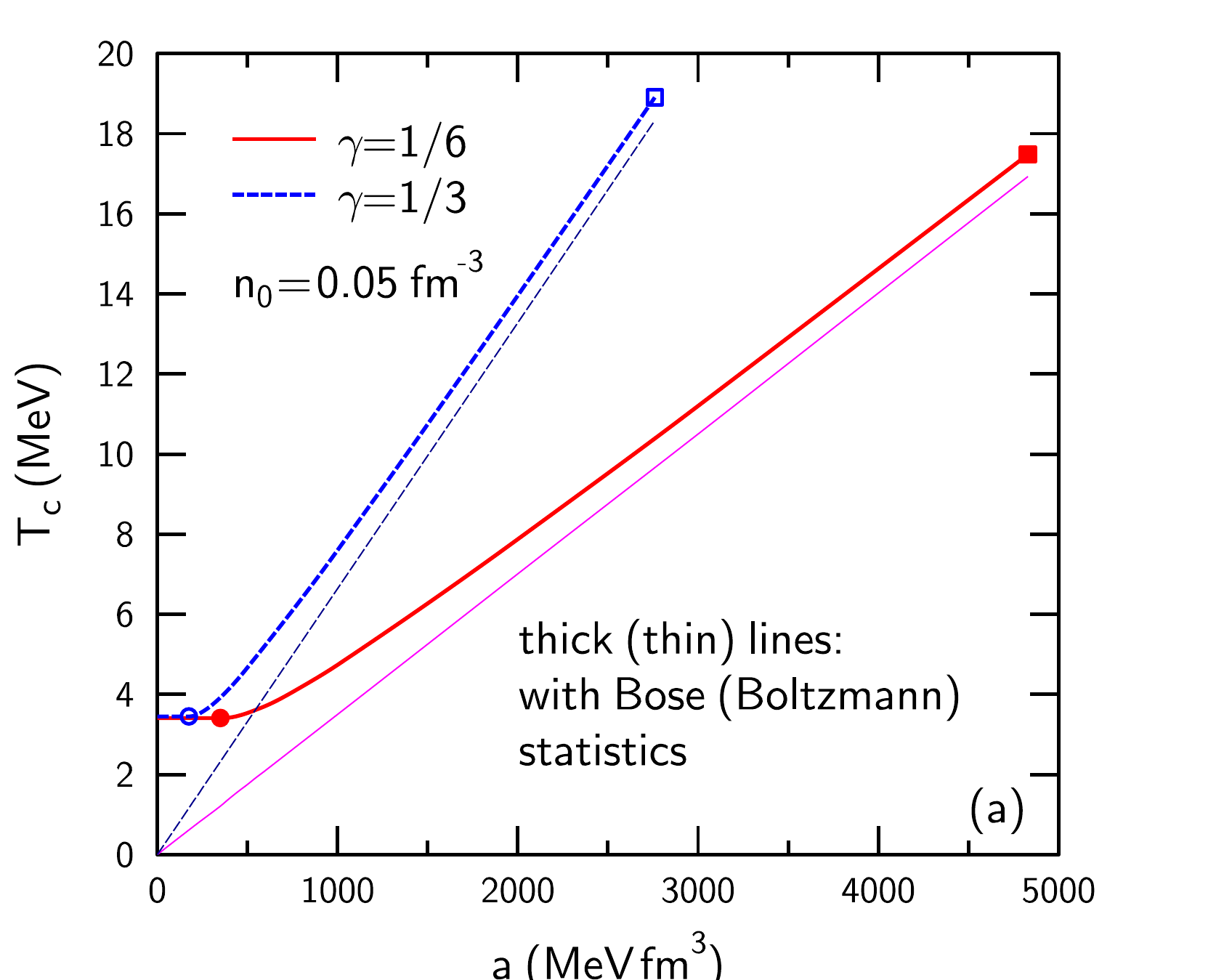}
\includegraphics[trim=2cm 7.5cm 2cm 8.5cm, clip, width=0.47\textwidth]{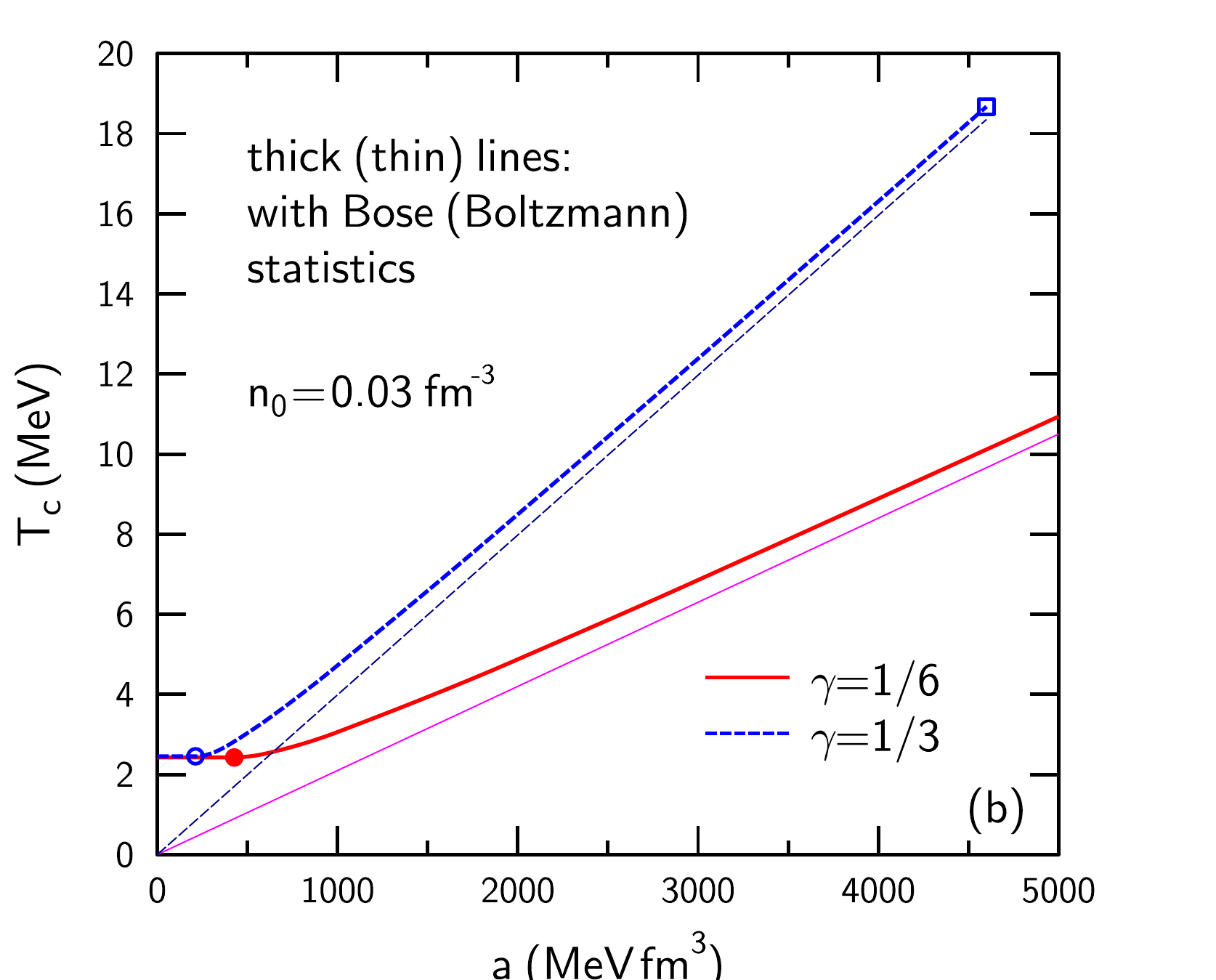}
\caption{
Critical temperature of $\alpha$-matter as a function of coupling
constant $a$ for $\gamma=1/6$ (solid curves) and $1/3$ (dashed lines).
Thick and thin lines correspond, respectively, to quantum-statistical
and Boltzmann calculations. The dots show the onset of
the saturation regime (see the text). The squares give maximum possible values
of the parameter $a$. (a): $n_0=0.05$~fm$^{-3}$ and (b):~\mbox{$n_0=0.03$~fm$^{-3}$}.
}\label{fig-4}
\end{figure}

The calculation shows that the Boltzmann approximation (\ref{crp1}) is
rather accurate at large values of~$a$, but it breaks down at small $a$ as shown in Fig.~\ref{fig-4}.
The CP can be found as the intersection of spinodal lines which go through the local minima and maxima of pressure isotherms. Let us consider points of the right-hand side (RHS) spinodal
in the $(n,T)$ plane as a function of temperature. At small~$T$,
the states on this spinodal can be found analytically by solving the equation
$(\partial p/\partial\hsp n)_T=0$.
Below the BEC line $\mu^*=m$, the first term in~(\ref{pf}) does not contribute
to $(\partial p/\partial\hsp n)_T$, and, therefore, the above equation is equivalent to
the condi\-tion~$p_f^{\hsp\prime}(n)=0$. From \re{pf1} we conclude that
the RHS spinodal at $T\leqslant T_{\rm BEC}(n^*)$ corresponds to a fixed density $n=n^*$:
\bel{n*}
n^*~=~n_0\,\left(\frac{2}{2+\gamma}\right)^{1/\gamma},
\ee
On the other hand, the left-hand side spinodal lies above the BEC line.
This proves that  $T_{\rm c}\geqslant T_{\rm BEC}(n^*)$
for all values of $a$.

At large enough values of the coupling constant $a$ the CP position is above the BEC line, i.e., the states at the RHS spinodal move further to the left ($n<n^*$) and
up\-wards~(\mbox{$T>T_{\rm BEC}(n^*)$}) in the $(n,T)$ plane.
In particular, this occurs, for the model parameters given in Table~I.
However, at small values of $a$ the situation is changed. One can calculate analytically the behavior
of~$(\partial p_{\rm id}/\partial\hsp n)_T$ at $(m-\mu ^*)\rightarrow 0$ \cite{BG}, i.e., in the vicinity of the BEC line. The calculation shows that if the condition
\bel{as}
a~\leqslant ~a_{\rm s}~=~\frac{\zeta^2(3/2)}{4\pi}~\frac{T_{\rm BEC}(n^*)}{\gamma\,n^*}
\ee
holds, then the equation $(\partial p/\partial n)_T =0$ has no solutions at $T>T_{\rm BEC}(n^*)$. In this case the~CP has a fixed position
\bel{T*}
T_{\rm c}=T_{\rm BEC}\hsp (n^*)\,,~~~~~ n_{\rm c}~=~n^*~~~~~~(a\leqslant a_{\rm s})\,.
\ee

In Figs.~\ref{fig-3} (b) and (d) we present the results for $\gamma=1/6$ and \mbox{$n_0=0.05~\textrm{fm}^{-3}$} (these parameters are the same as in panels (a) and (c)), but choose $a=0.2$~GeV~fm$^3$. In this case,~$a$ is smaller
than the value $a_{\rm s}\simeq 358$~MeV~fm$^3$~obtained from (\ref{as})
and, therefore, the~CP position is determined by~\re{T*}.
Several pressure isotherms are shown in Fig.~\ref{fig-3}~(d). Thin lines
represent metastable and unstable parts of the isotherms. The position of the RHS spinodal
is shown by the vertical dash-dotted line. One can see a 'kink' between the left- and right-hand
side binodals at the CP.

\begin{table}[ht!]
\caption
{\footnotesize Characteristics of the CP in the saturation regime.}
\label{tab2}
\vspace*{3mm}
\begin{tabular}{|c|c|c|c|c|c}\hline
~$n_0$~(fm$^{-3})$ &~~~$\gamma $~~~&~$a_{\rm s}~(\textrm{MeV~fm}^3)$~&~
$n^*~(\textrm{fm}^{-3})$~&~$T_{\rm BEC}\hsp (n^*)~(\textrm{MeV})$~\\
\hline
0.05~~& 1/6& 358 & 0.0310 & 3.41  \\
\hline
0.05~~&1/3 & 178 & 0.0315 & 3.45  \\
\hline
0.03~~&1/6 & 427 & 0.0186 & 2.43   \\
\hline
0.03~~&1/3 & 212 & 0.0189 & 2.46  \\
\hline
\end{tabular}
\end{table}

The dependence of $T_{\rm c}$ on the parameters $a,~\gamma,~n_0$
is illustrated in Fig.~\ref{fig-4} and in Table~II.
The upper limits for $a$ in Fig.~\ref{fig-4} are taken from the requirement
that the binding energy per baryon in the ground state of $\alpha$-matter is smaller
than 16~MeV, i.e., the estimated binding energy
of homogeneous isospin-symmetric nucleon matter at $T=0$.
As seen from Fig.~\ref{fig-4} the Boltzmann approximation
fails at small values of $a$.
Therefore, the saturation of $T_{\rm c}$
at~$a\rightarrow 0$ is a pure quantum effect, a consequence
of the Bose statistics. Due to the BEC effects, the system exhibits the LGPT
with the CP given by Eq.~(\ref{T*}) even for an infinitely small interaction coupling $a$.
This happens due to the special form (\ref{skyrme}) of the interaction potential $U(n)$, but may occur also in other situations.

Some properties of the CP at $a\leqslant a_{\rm s}$ are similar to those at $a>a_{\rm s}$:
1) it is the end point of the LGPT; 2) $(\partial p/\partial\hspm n)_T=0$ and $\omega=\infty$ at
$T=T_{\rm c}$ and $n=n_{\rm c}$. However, there are some essential differences: at $a>a_{\rm s}$, one has
in addition $(\partial^{\hsp 2}p/\partial\hspm n^2)_T=0$. Hence, the behavior of thermodynamical functions in the vicinity of the CP is governed by the critical indices which belong to the universality
class of the van der Waals model (the so-called mean-field theory). At $a\leqslant a_{\rm s}$,
the second derivative $(\partial^{\hsp 2}p/\partial\hsp n^2)_T$ has a discontinuity at the CP.
In this case the standard concept
of the critical indices should be reconsidered. Note that~in this case
the left- and right-hand side binodals are essentially different in the vicinity of the CP: the left-hand side binodal consists
of gas states with $n_{\rm bc}=0$, whereas the states on right one correspond to the liquid with a~nonzero Bose-conden\-sate.

\section{Concluding remarks}

We have analyzed the phase diagram of pure $\alpha$-matter within a simple model
using a~density-dependent mean-field interaction.
This model does not only predict the liquid-gas
mixed phase at low temperatures, but simultaneously it describes the Bose-Einstein condensation. The interplay between these two phenomena shows new important effects.

The end point of the LGPT is located on the BEC line at small values of the interaction coupling, where
the standard description of the critical point is
not applicable. This behavior may also occur in other multiparticle systems.
In particular, we expect to find similar effect in atomic Bose systems.

Pure $\alpha$-matter at nonzero temperatures
is an idealization which does not respect the chemically equilibrium
conditions. A more realistic approach should include also nucleons and nucleon
clusters like $d,t,^3$He. The equilibrium mixture of nucleons and light nuclei are
important for heavy-ion collisions and astrophysical applications.
This topic will be addressed in future studies.

\begin{acknowledgments}
The authors thank D.~V.~Anchishkin for useful discussions.
A.M. is thankful for the support from the Eurasia grant No.~CPEA-LT-2016/10094.
The work of M.I.G. is supported by the
Goal-Oriented Program of  the National Academy of Sciences of Ukraine, by the
European Organization for Nuclear Research (CERN), Grant CO-1-3-2016,
and by the Program of Fundamental Research of the Department of Physics and
Astronomy of National Academy of Sciences of~Ukraine. V.V. appreciates
the support from HGS-HIRe for FAIR.
\end{acknowledgments}

\end{document}